\documentclass[oldversion]{aa}
\usepackage{amsmath,graphicx,amssymb}
\usepackage{txfonts}
\usepackage{natbib}
\bibpunct{(}{)}{;}{a}{}{,}

\newcommand{\hzav}[1]{\left[#1\right]}
\newcommand{\oc}{\textit{O-C}}

\begin{document}

\title{Kwee-van Woerden method: To use or not to~use?}


\author{
        Z.\,Mikul\'a\v{s}ek \inst{1,2}
      \and
        M.\,Chrastina \inst{1}
      \and
        J.\,Li\v{s}ka \inst{1}
      \and
        M.\,Zejda \inst{1}
        \and
        J.\,Jan\'{i}k \inst{1}
        \and
        L.-Y.\,Zhu \inst{3,4}
        \and
        S.-B.\,Qian \inst{3,4}
       }

\institute{
            Department of Theoretical Physics and Astrophysics, Masaryk University, Brno, Kotl\'a\v{r}sk\'{a} 2, CZ-611\,37 Brno, Czech Republic
         \and
            Observatory and Planetarium of Johann Palisa, V\v{S}B--Technical University, Ostrava, Czech Republic
         \and
            National Astronomical Observatories/Yunnan Observatory, Chinese Academy of Sciences, Kunming, China
         \and
            Key Laboratory for the Structure and Evolution of Celestial Objects, Chinese Academy of Sciences, Kunming, China
          }

\date {Received}

\abstract {The trustworthiness of orbital period analyses of eclipsing binaries strictly depends on the correctness of the observed mid-eclipse time determination, as well the reliability of its uncertainty estimation. The majority of them has been determined by means of the Kwee-van Woerden method (KWM). There are also other possibilities – e.g. to use physical models of eclipsing binaries or light curve templates and to determine mid-eclipse times using the least square method (LSM). We compared results yielded by both methods by means of a computer simulation on the synthetic model of AR\,Aur primary minimum. Minima times determined by the KWM and the exact LSM approach are nearly the same, while the scatter of LSM times is always smaller than the scatter of KWM times. KWM uncertainties are systematically underestimated. We think that the time is ripe for the Kwee-van Woerden method to retire.}


\titlerunning{Kwee-van Woerden method: To use or not to use?}
\authorrunning{Z.~Mikul\'a\v{s}ek et al.}
\maketitle

\section{Motivation. Kwee-van Woerden and LSM methods}\label{mot}

Exploration of eclipsing binaries (EB) by their O-C diagrams is one
of the most powerful instruments of modern stellar astrophysics. It serves as a sophisticated period analyses tool that can reveal a number of intimate details of binary stars' lives. The trustworthiness of orbital period analyses of eclipsing binaries strictly depends on the correctness of the observed mid-eclipse time determination, as well the reliability of its uncertainty estimation.

A well-known shortcoming of practically all eclipsing binary \oc\ diagrams is that the scatter of individual $(\oc)_i$  values is many times larger than the scatter expected from their uncertainties $\delta O_i$. Mathematically:
\begin{equation}
\chi^2_{\mathrm{r}}\cong\frac{ \chi^2}{N}=\frac{1}{N}\sum_{i=1}^N\hzav{ \frac{(\oc)_i}{\delta O_i}}^2\gg1,
\end{equation}
where $N$ is the number of individual times of minima.

\begin{figure}
\centering \resizebox{0.99\hsize}{!}{\includegraphics{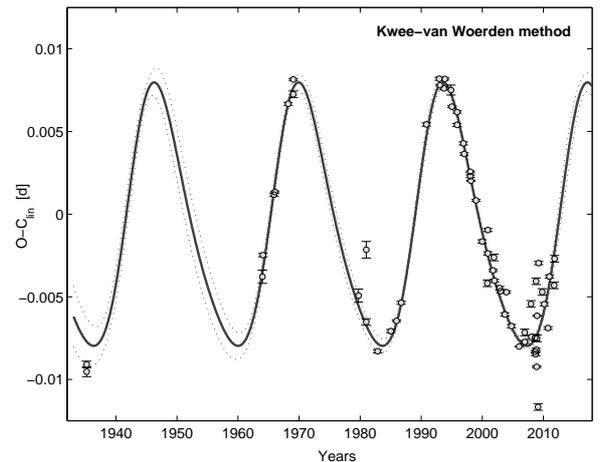}}
\caption{\small{The \oc\ diagram of AR Aurigae with mid-eclipse times and uncertainties derived from original data using the Kwee-van Woerden method.}} \label{kwoc}
\end{figure}

This can be illustrated by means of an example of a detached eclipsing binary, AR Aur, with a well-defined light time effect. We collected all available original photoelectric measurements obtained during 65 primary/secondary minima and derived from them mid-eclipse times ${O_i}$ including their uncertainties $\delta O_i$. At the same time we calculated the predicted times of corresponding minima according to the ephemeris taking into account the influence of the third body in the system (see fig.\,\ref{kwoc}).  We found that $\chi^2_{\mathrm{r}}=158$!

All mid-eclipse times and their uncertainties were derived by the unique standard procedure based on the original \citet{kwee} method commonly used for such purposes. There are two possible explanations for our poor results: a) uncertainties $\delta O_i$ could be greatly underestimated; b) the method of mid-eclipse time determination gave incorrect results. With respect to magnitude of $\chi^2_{\mathrm{r}}$ it is very likely that we here encountered the combination of both possibilities.

The majority of published EB minimum times $O$ and their uncertainties $\delta O$ were determined by means of the famous Kwee-van Woerden method from 1956. The methods has many advantages: it yields indisputable results, it requires no assumptions about the form of LC shapes (except their symmetry), it needs only very simple computational techniques. Consequently, K-W method is generally accepted and used by most EB astronomers all over the world.

Nowadays, the majority of the KWM users do not use the original version of the method, but some modifications hidden in the code packages. These codes then act as KWM black boxes that are able to calculate the demanded result for an observational time series {ti;mi} - the minimum $\{t_i,m_i\}$ to calculate the demanded result - the minimum time $O$ and its error $\delta O$. Only the authors of these KWM codes may know the true content of their PC black boxes.

Nevertheless, there is also another possibility for getting the result, e.g. to use physical or phenomenological models of LC templates  and to determine $O$ and its error $\delta O$ using the transparent least square method.

\section{Simulations}

The aim of this short paper is to compare the results of both methods by means of computer simulations. The object of simulations was the modelling of the $V$ light curve of AR Aur during its primary minimum scattered by random numbers with normal distribution. The KWM code was written according to the original K--W paper\footnote{We found that the original KWM is usable only for light curves sparsely populated by measurements with relatively small and moderate scatter. It cannot be used for standard CCD observations. Fortunately, after some LSM preprocessing of data we were able to follow the scheme of original KWM.}. The template light curves for the LSM method was the non-scattered models of the LC. We did 500\,000 simulations with $\sigma=$ 0.0025–-0.0325 mag and with durations of observations from 2.4 to 5.6 h for three situations: a)~the strictly symmetric light curve without any trend with the minimum in the middle, b)~the symmetric LC influenced by the typical trend of d$m/$d$t=0.005$\,mag\,h$^{-1}$  with the minimum in the middle, and c)~the symmetric LC without any trend with an asymmetrically (2:3) placed minimum.

\section{Conclusions}

\begin{figure}
\centering \resizebox{0.836\hsize}{!}{\includegraphics{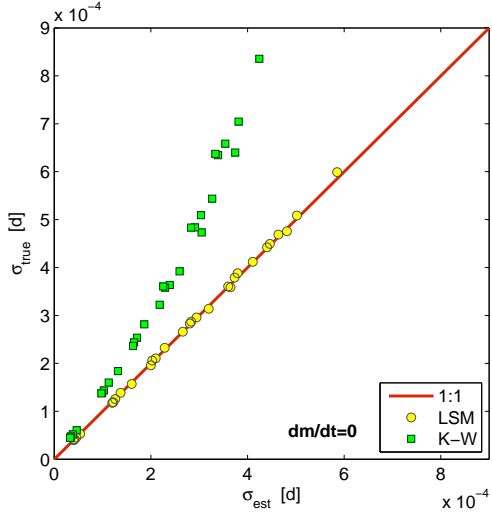}}
\caption{\small{Dependence between the estimated and true scatters for LSM and KWM in the presence of the trend clearly shows that LSM estimates are correct, while KWM ones are always strongly underestimated.}} \label{unoc}
\end{figure}

\begin{figure}
\centering \resizebox{0.99\hsize}{!}{\includegraphics{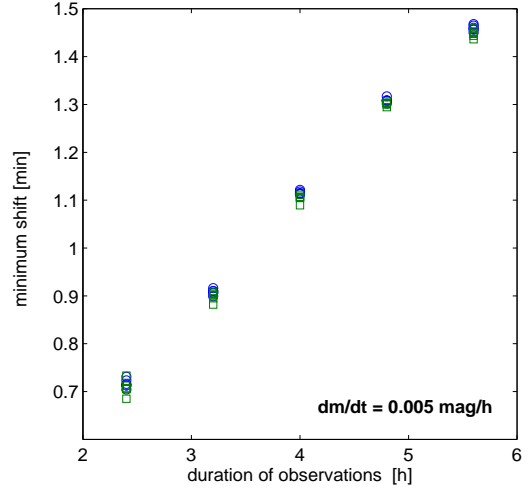}}
\caption{\small{Unreduced trends during nights equally influence mid-eclipse times determined by both methods. The dependence shows the necessity to use such models which are able to implement trend corrections. KWM - squares, LSM - circles.}} \label{shidur}
\end{figure}

We found that minima times determined by the KWM and the exact LSM approach are nearly the same. Consequently, the published KWM times of minima can be used as good estimates. The scatter of LSM times is always smaller than the scatter of KWM times. The additional information hidden in knowledge of the LC shape is a bonus.

While the uncertainty estimations of LSM times are nearly realistic, estimations of the uncertainty of KWM times are always underestimated by a factor of 1.4 in the case of no trend. A much larger discrepancy occurs when the trend does exist.

Both methods are equally influenced by uneliminated trends common for differential photometry in CCD observations. The shift rises with the duration of observations. A~typical trend of d$m/$d$t=0.005$\,mag\,h$^{-1}$ (only linear term is decisive) causes a shift from 0.7 to 1.6 min of minima timings, depending on duration (see Fig.\,\ref{shidur}).

The LSM approach enables correction of mid-eclipse times for linear trends. The correction diminishes the scatter of \oc\ values three times. At the same time the uncertainties of $O$ times roughly doubles (more observing intervals). Both effects suppress $\chi^2_{\mathrm{r}}$ from an alarming 158 to an also bad, but more acceptable 12. The KWM also zields results that are greatly inferior in the case of incomplete LCs or asymmetrically placed minima.

Should the much favoured Kwee-van Woerden method for variable star data processing be used? We think the time for the KWM to be abandoned is at hand.

\begin{figure}
\centering \resizebox{0.99\hsize}{!}{\includegraphics{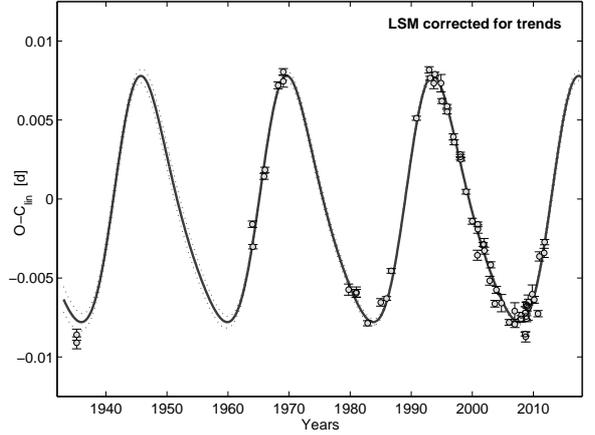}}
\caption{\small{The \oc\ diagram of AR Aurigae with mid-eclipse times and uncertainties derived from original data by LSM corrected for possible trends during nights (compare with fig.\,\ref{kwoc})}.} \label{lsmoc}
\end{figure}

\acknowledgements
The investigation was partly supported by the project LH12175. Authors thank S. de Villiers for the valuable discussion and careful language revision.


\begin{thebibliography}{}

\bibitem[Kwee \& van Woerden(1956)]{kwee} Kwee, K.K., van Woerden, H. 1956, \textit{Bull. Astron. Inst. Neth.}, 9, 252

\end{thebibliography}
\end{document}